\documentclass[runningheads]{llncs}
\usepackage{graphicx}
\usepackage{multirow}
\usepackage{subcaption}
\usepackage{url}

\begin{document}
\title{Uncertainty-based quality assurance of carotid artery wall segmentation in black-blood MRI}
\titlerunning{Uncertainty-based quality assurance of carotid artery segmentation}
%
\author{Elina Thibeau-Sutre \and
Dieuwertje Alblas \and
Sophie Buurman \and 
Christoph Brune \and
Jelmer M. Wolterink}
\authorrunning{E. Thibeau-Sutre et al.}
%
\institute{Mathematics of Imaging and AI, Department of Applied Mathematics, Technical Medical Centre, University of Twente, Enschede, The Netherlands
\email{\{e.thibeau-sutre,j.m.wolterink\}@utwente.nl}}
\maketitle              
\begin{abstract}
The application of deep learning models to large-scale data sets requires means for automatic quality assurance. We have previously developed a fully automatic algorithm for carotid artery wall segmentation in black-blood MRI that we aim to apply to large-scale data sets. This method identifies nested artery walls in 3D patches centered on the carotid artery. 
In this study, we investigate to what extent the uncertainty in the model predictions for the contour location can serve as a surrogate for error detection and, consequently, automatic quality assurance. 
We express the quality of automatic segmentations using the Dice similarity coefficient. The uncertainty in the model's prediction is estimated using either Monte Carlo dropout or test-time data augmentation. 
We found that (1) including uncertainty measurements did not degrade the quality of the segmentations, (2) uncertainty metrics provide a good proxy of the quality of our contours if the center found during the first step is enclosed in the lumen of the carotid artery and (3) they could be used to detect low-quality segmentations at the participant level. This automatic quality assurance tool might enable the application of our model in large-scale data sets.
\end{abstract}

\section{Introduction}
Ischemic stroke refers to a neurological deficit due to insufficient blood flow to the brain and may lead to long-term disability \cite{laiPersistingConsequencesStroke2002}. Its prevalence is expected to rise in the coming years in the European population, leading to the need for improved prevention of this medical condition \cite{wafaBurdenStrokeEurope2020}. Imaging of the head and neck can provide information on vascular biomarkers, such as the geometry and anatomy of the carotid arteries, which provide the brain with blood \cite{baluCarotidPlaqueAssessment2011}.  For example, it has been shown the thickness of the carotid artery wall and its geometry are risk factors for stroke independently from traditional risk factors such as hypertension \cite{phanCarotidArteryAnatomy2012,chamblessCarotidWallThickness2000}. 

Unraveling the relationship between vascular biomarkers and stroke requires large-scale imaging studies, in which accurate manual segmentation of relevant structures is a tedious task. To alleviate this problem, there have been efforts to develop deep learning algorithms for carotid artery segmentation in MRI~\cite{wangApplicationArtificialIntelligence2023,zieglerAutomatedSegmentationIndividual2021,huang3DCarotidArtery2023}. One of these methods included an uncertainty regularisation component in their loss and used uncertainty to interpret their network \cite{lavrovaURCarANetCascadedFramework2023}. Likewise, we have developed a deep learning-based algorithm for carotid artery inner and outer wall segmentation that exploits symmetry and anatomical priors and achieved top-ranking quantitative and qualitative results in the MICCAI \& SMRA 2021 Carotid Artery Vessel Wall Segmentation Challenge~\cite{alblasDeeplearningbasedCarotidArtery2022}.



While our method achieved excellent results on a curated challenge dataset, it is likely to make errors when applied to large-scale data sets. Automatic identification of such errors using a quality assurance mechanism would have major practical value. 
Here, we study the feasibility of developing such a system by investigating the use of uncertainty quantification of the algorithm as a proxy for the quality of the segmentations it produces. We distinguish two types of uncertainty used in Bayesian modelling: the \textit{aleatoric uncertainty} that can be attributed to noise in the data, and the \textit{epistemic uncertainty}, which represents the limitation of the model and its ability to capture the underlying distribution of the data \cite{kendallWhatUncertaintiesWe2017}. Correlations between these different types of uncertainties and the quality of deep learning predictions were previously found in medical image analysis \cite{wangAleatoricUncertaintyEstimation2019,royBayesianQuickNATModel2019}. We here study this relation in the context of carotid artery segmentation in black-blood MRI where we focus on two potential sources of error: variations in image quality, and variations in the quality of a pre-processing step that is essential to our algorithm. 

\section{Methods}

\subsection{Data set}
\label{sec:data}
Experiments were conducted using the black-blood MRI training set of the MICCAI \& SMRA 2021 Carotid Artery Vessel Wall Segmentation Challenge~\cite{yuanCarotidVesselWall2021}. This data set is a subset of the larger CARE-II data set, acquired from 13 different hospitals and medical centers throughout China~\cite{zhaoChineseAtherosclerosisRisk2017}. In the training subset, images were acquired according to two different protocols, leading to two different distributions (20 patients in the first distribution and three in the second one). Only one black-blood MRI image is available per participant. Manual annotations of the contour of the lumen and outer wall of the common, internal, and external carotid artery were provided by the challenge organizers in a subset of axial image slices, for both the left and the right carotid arteries. We preprocessed each image volume by rescaling its intensities between 0 and 1 according to its 5th and 95th intensity percentiles. All volumes were oriented according to the RAS coordinate system.

\begin{figure}[t!]
\centering
\includegraphics[width=\textwidth]{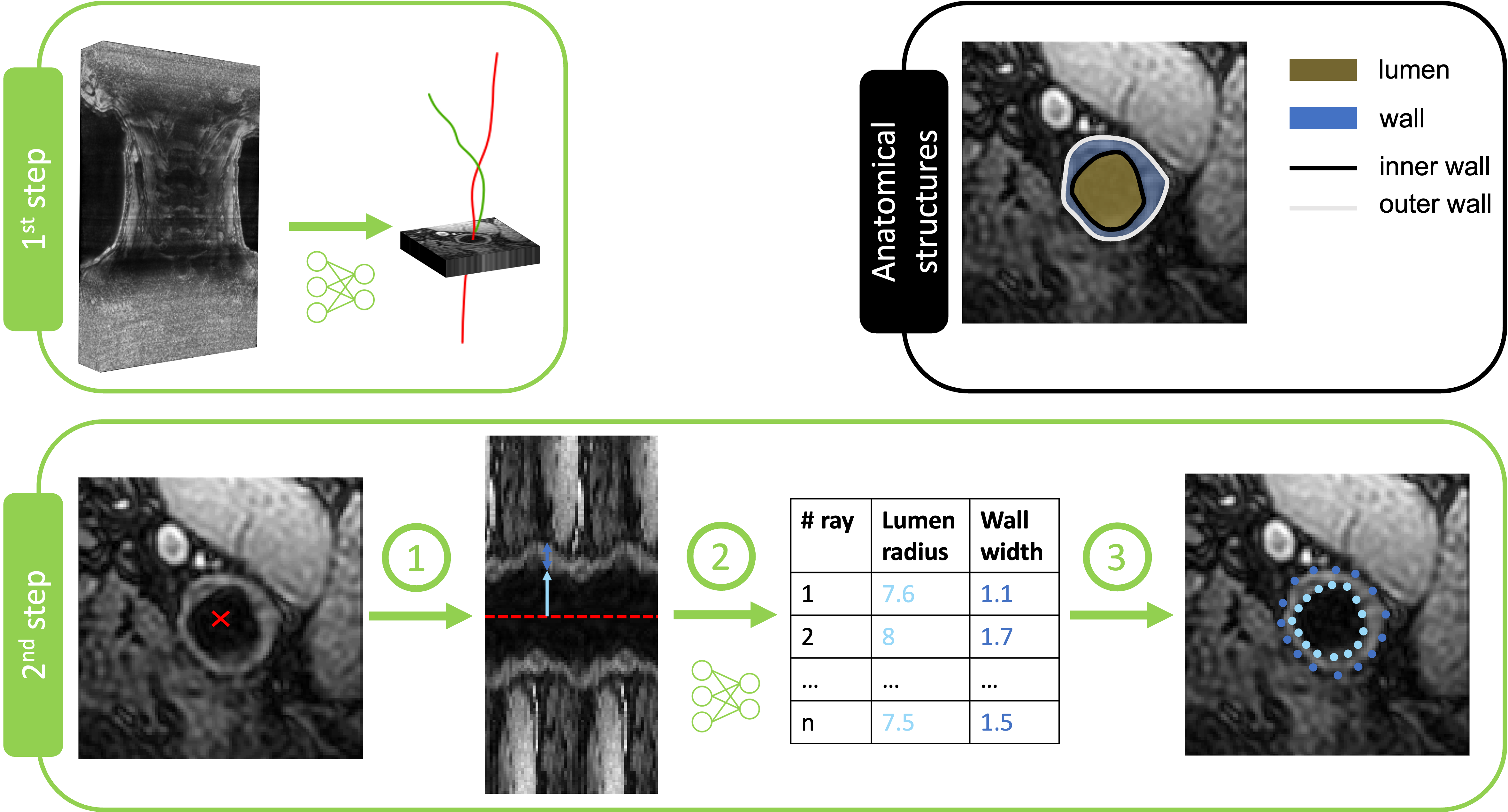}
\caption{Illustration of our previously developed automatic segmentation pipeline. The first step consists of finding the centerlines of the internal (red) and external (green) carotid arteries. The second step consists of (1) a polar coordinate transformation, (2) regression of the lumen radii (light blue) and wall widths (dark blue), and (3) inversion of the polar transform~\cite{alblasDeeplearningbasedCarotidArtery2022}} \label{fig:pipeline}
\end{figure}

\subsection{Segmentation method}
We previously developed a deep learning-based algorithm for the segmentation of carotid black-blood MRI images~\cite{alblasDeeplearningbasedCarotidArtery2022}. This method consists of two steps. In the first step, the centerlines of the left and right common, internal, and external carotid arteries are detected. For each artery centerline, a 3D U-Net predicts a proximity map, which is subsequently used for centerline extraction using Dijkstra's algorithm. For both the internal and the external carotid artery, we find a centerline that includes the common carotid artery. Hence, for each 3D volume, we find four centerlines. 

In the second step, contours for the inner and outer wall are detected for each centerline point. We exploit rotation equivariance in the data by transforming our Cartesian images to a polar coordinate system centered at the centerline point, in three steps:
\begin{enumerate}
    \item \textbf{Polar transform}: We cast 31 equiangular rays of length 127 pixels from the center in the axial plane. This is also done for the three adjacent slices in both directions so that we obtain a cylindrical volume of 31x127x7 voxels.
    \item \textbf{Regression of radii and widths}: For each of the rays in the center slice, and hence for each angle, a translation equivariant 3D CNN estimates the radius of the lumen wall and the nonzero offset between the lumen wall and the outer wall. By adding this offset to the lumen wall radius, we obtain the outer wall radius for each angle, resulting in a 31x2 matrix of predictions.
    \item \textbf{Polar transform inversion}: We transform the detected contours back to the original Cartesian space and obtain a ring-shaped segmentation of the vessel wall.
\end{enumerate}
This method provides closed, non-intersecting contours for the lumen wall and outer wall for each centerline point, which can be combined into watertight meshes. In the MICCAI \& SMRA 2021 Carotid Artery Wall Segmentation Challenge, this led to results that were both quantitatively and qualitatively superior to all other submissions. An implementation of this method is available online\footnote{\url{https://github.com/MIAGroupUT/carotid-segmentation}}.

The true center corresponds to the center of the manually annotated contour of the lumen. However, in practice as the center is estimated automatically during the first step, it is not always perfectly centered, and in the worst case located outside the vessel lumen. In that case, the segmentation will fail, as we are unable to perform the transformation to a polar coordinate system, and the assumption that the lumen wall radius and offset to the outer wall add up no longer holds. Moreover, image quality in black-blood MRI can vary widely, leading to uncertainty in the estimated lumen radius and wall thickness. Hence, in this study, we aim to assess the robustness and uncertainty of the procedure toward the placement of the center, as well as the image quality.


\subsection{Simulation of inputs}
To simulate what our model might encounter in large-scale datasets, we simulate two types of data using the original data set described in Sec. \ref{sec:data}. First, we reduce the image quality by linearly adding Gaussian noise to the original image $\alpha * noise + (1 - \alpha) * image$, where $\alpha$ is a noise level. 

Second, we simulate the effect of poor-quality centerline extraction in the first step of our algorithm. 
To reproduce this issue, centers with different spatial offsets were generated. Based on the manual contour annotations in training samples, we define the true center of the artery in a cross-sectional image. The point on the inner wall contour with the largest distance to the true center is taken to find the direction along which new centers are sampled. The spatial offset corresponds to the distance to the true center normalised by the radius. Consequently, a spatial offset of 0 corresponds to the true center, an offset between 0 and 1 to a center inside the lumen, and an offset above 1 to a center outside the lumen. This last option corresponds to the scenario in which an improper centerline was found in the first step of the algorithm.

\subsection{Training and evaluation of the networks}
The CNNs assessing the lumen radii and wall widths were trained using the open-source implementation of our method.
The CNNs have eight convolutional layers, between which Dropout layers with a rate of 0.2 are included. The training loss is the mean squared error on the lumen radii and wall widths. The model was trained for 50 epochs, and the final version corresponds to the one which obtained the best validation loss during training at the end of an epoch during training.

\subsection{Uncertainty quantification and quality assessment}
In this study, we aim to investigate the correlation between segmentation quality and uncertainty estimation.  The quality of the segmentations is assessed using the Dice similarity coefficient between the manually annotated and automatically computed contours. 
To estimate the uncertainty of the model, two different methods were implemented:
\begin{enumerate}
    \item $dropout$ corresponds to Monte Carlo Dropout, in which we keep the Dropout layers active during inference and estimate the contours 20 times. This procedure was introduced to approximate Bayesian inference and computes an estimation of epistemic uncertainty \cite{galBayesianConvolutionalNeural2016}.
    \item $centers$ computes contours based on polar transformed images acquired using the eight neighboring pixels of the original centers. This can be seen as a form of test-time augmentation, a procedure for improved aleatoric uncertainty estimation \cite{ayhanExpertvalidatedEstimationDiagnostic2020}.
\end{enumerate}

Subsequently, the variability on the result is estimated using two different methods:
\begin{enumerate}
    \item $mean$ computes for each ray the mean value and the standard deviation of the distances, corresponding to an output distance and its uncertainty, respectively.
    \item $polar$ computes the polar coordinates of all the points of the set of contours and fits a degree 2 polynomial model between the sine and cosine of the angle and the distance. The output distance corresponding to an angle $\theta$ is computed using this model, and the uncertainty corresponds to the distance to the model of all the points in $[\theta - \frac{1}{32}\pi; \theta + \frac{1}{32}\pi]$ (see supplementary Figure 1).
\end{enumerate}
The $polar$ method was developed to estimate the final contours and uncertainties using the $centers$ variability, and it can also be used with $dropout$. In both cases, the result is normalised by the output distance, as we observed that this led to better results.

The correlation between the uncertainty and the Dice score was assessed using linear regression and quantified using the coefficient of determination $R^2$. This correlation was computed at three different levels. At the contour level, the uncertainties of the 31 points of a contour were averaged and correlated to the Dice score of the corresponding contour. At the vessel or participant level, the uncertainties of the points of all annotated contours were averaged and compared to the average Dice score of all contours of one vessel or one participant, respectively. The lumen and wall contours were considered separately to compute the correlation strengths. Illustrations are given in Appendix A1 (Figure 2).

\section{Results}

The three images acquired with a different protocol were put in the test set to assess if the correlation between uncertainty estimation and image quality holds on a distribution never seen by the network. Among the other 20 images, ten were put on the test set and ten were in the training/validation set of the networks. This last set is split into the training (eight images) and validation (two images) sets.

The Dice scores obtained without uncertainty estimation (0.78 $\pm$ 0.18) were similar to or less than those obtained with the different uncertainty estimations: $dropout_{mean}$ (0.82 $\pm$ 0.12), $dropout_{polar}$ (0.82 $\pm$ 0.12), $centers_{polar}$ (0.78 $\pm$ 0.18). To note, the three participants with different acquisition parameters obtained similar Dice scores as the others. 

\subsubsection{Image quality}

\begin{figure}[!htp]
\centering
\begin{subfigure}[b]{0.9\textwidth}
    \includegraphics[width=\textwidth]{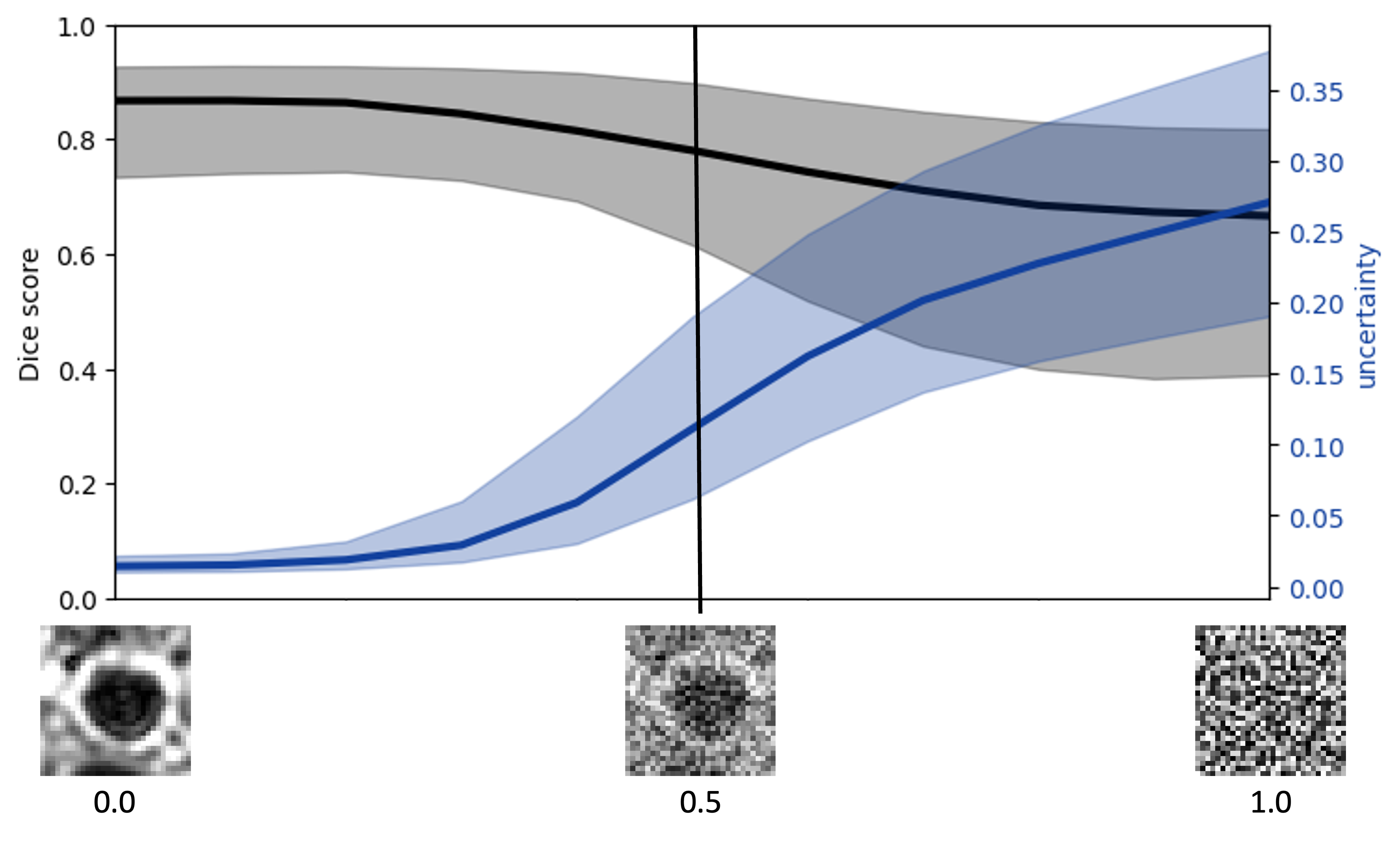}
    \caption{Image quality} 
    \label{fig:noise_quality}
\end{subfigure}
\begin{subfigure}[b]{0.9\textwidth}  
    \includegraphics[width=\textwidth]{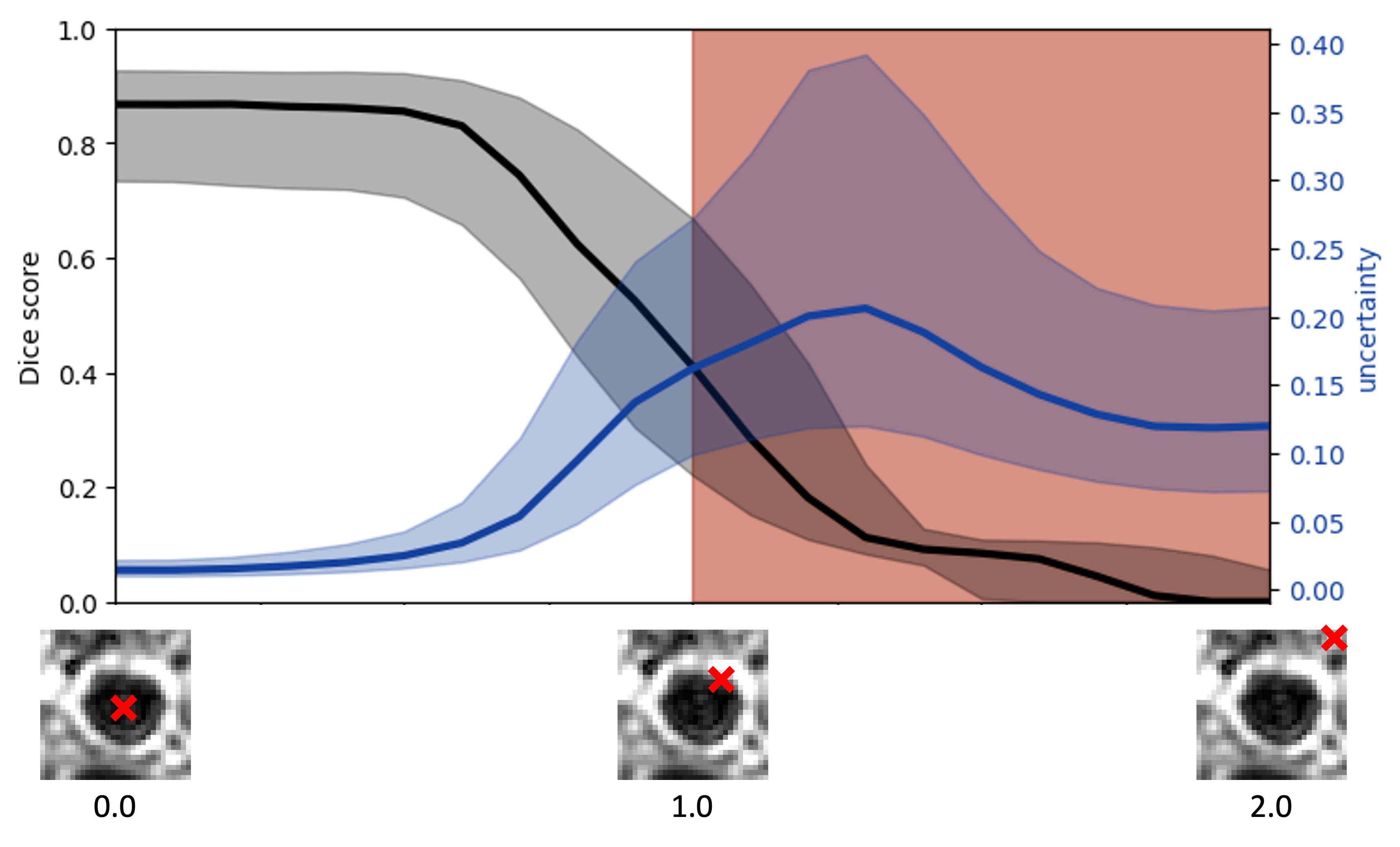}
    \caption{Center quality} 
    \label{fig:center_quality}
\end{subfigure}
\caption{Evolution of the quality (black) and the uncertainty (blue) while degrading the image quality (a) or the center quality (b). The colored space around the lines correspond to the inter-quartile range. The red region corresponds to centers which are outside the lumen.}
\label{fig:input_quality}
\end{figure}

Figure \ref{fig:noise_quality} shows that the quality of segmentations decreases as the level of noise increases. We observe a strong correlation between the quality of the image and the uncertainty of the segmentations at the participant level but not at the contour or vessel levels (Table \ref{tab:correlation}). The strongest correlations are obtained with $dropout_{polar}$. Note that the segmentation quality remains relatively high (mean Dice $>$ 0.6), as the true center is used to create the polar transformed images, and any correctly centered contour with reasonable radius estimates is bound to have a decent overlap with the annotation.

\begin{table}[!htp]
\caption{Values of the coefficient of determination $R^2$ of the correlation of the uncertainty and quality of the segmentations obtained with different levels of noise different centers in the lumen.}\label{tab:correlation}
\centering
\begin{tabular}{p{2cm} p{2cm} p{2cm}|c c c c }
\hline
Experiment & Correlation level & Structure & $dropout_{polar}$ & $dropout_{mean}$ & $centers_{polar}$\\
\hline
\multirow{6}{4em}{Image quality} & \multirow{2}{4em}{contour} & lumen & \textbf{0.33} & 0.31 & 0.24 \\
& & wall & \textbf{0.38} & 0.30 & 0.24 \\
\cline{2-6}
&\multirow{2}{4em}{vessel} & lumen & \textbf{0.59} & 0.56 & 0.29 \\
& & wall & \textbf{0.67} & 0.56 & 0.30 \\
\cline{2-6}
&\multirow{2}{4em}{participant} & lumen & 0.85 & \textbf{0.87} & 0.77 \\
& & wall & \textbf{0.88} & 0.83 & 0.72 \\
\hline
\hline
\multirow{6}{4em}{Center quality} & \multirow{2}{4em}{contour} & lumen & \textbf{0.56} & 0.50 & 0.54 \\
& & wall & \textbf{0.53} & 0.24 & 0.42 \\
\cline{2-6}
&\multirow{2}{4em}{vessel} & lumen & \textbf{0.81} & 0.75 & 0.80 \\
& & wall & \textbf{0.85} & 0.73 & 0.64 \\
\cline{2-6}
&\multirow{2}{4em}{participant} & lumen & 0.93 & 0.89 & \textbf{0.96} \\
& & wall & \textbf{0.95} & 0.86 & 0.90 \\
 \hline
\end{tabular}
\end{table}

\subsubsection{Center quality}
Figure \ref{fig:center_quality} shows that the uncertainty of the segmentation model increases while the Dice score decreases until a spatial offset of 1.3 is reached. After that, the uncertainty decreases again though the quality is still decreasing. We assessed the correlation between uncertainty and quality for spatial offsets $<$ 1, which corresponds to centers in the lumen.
We observe a strong correlation between the quality of the segmentations and the uncertainty of the model at the participant and vessel levels, but still not at the contour level (Table \ref{tab:correlation}), obtained with $dropout_{polar}$.

\section{Discussion and conclusion}
We find that across different potential sources of error in our segmentations, the estimated uncertainty in the model's predictions is a good proxy of the quality of the segmentation obtained. This correlation is particularly present at the participant level. Interestingly, we found that for shifting artery centers, the correlation was only present when the centerline point was still inside the artery. This is the case for 95\% of slices in the data set (see Appendix A2). 

Depending on the context, the best correlation between segmentation quality and uncertainty estimate was obtained with a different method. In most cases, $dropout_{polar}$ performs best, and the second best is $centers_{polar}$ to assess the center quality, whereas $dropout_{mean}$ better estimates the image quality. This confirms the use of Monte Carlo Dropout as a robust uncertainty estimator. In practice, a combination of uncertainty estimation methods could be considered to improve further the correlation with segmentation quality, and thus the value of uncertainty estimation as a means to quality assurance.


A limitation of this study is the set of contours that were manually annotated: the bifurcation between the external and internal carotids, which is the most difficult region to segment for our algorithm, was only scarcely annotated by the organisers of the challenge. Annotations that we will collect in new data sets to assess the performance of our algorithm should focus on this region. Future work will further focus on the automatic assessment of the quality of the first step in our algorithm to build our final quality check tool that will be applied to large data sets.

In conclusion, we have shown that model uncertainty can serve as a proxy for segmentation, with the potential to provide automatic quality assurance of our model in large-scale data sets.

%

\bibliographystyle{splncs04}
\bibliography{main}

\end{document}